\documentclass[12pt]{article}

 \newcommand {\nc}{\newcommand}

 \nc{\be}{\begin{equation}}

 \nc{\ee}{\end{equation}}

 \nc{\bea}{\begin{eqnarray}}

 \nc{\eea}{\end{eqnarray}}

 \def\complex{{\cal C}}

 \def\integer{{\cal Z}}

 \def\natural{{\cal N}}

 \def\unit{{\bf 1}}

 \def\tld{\tilde}

 \def\bra{\langle}

 \def\ket{\rangle}

 \title{\bf Quiver Mechanics for Deconstructed\\
       Matrix String}

 \author{
  Jian Dai
  \thanks{jdai@physics.utah.edu},
  Yong-Shi Wu
  \thanks{wu@physics.utah.edu}\\
  Department of Physics, University of Utah\\
  Salt Lake City, Utah 84112}

\begin{document}

  \maketitle

  \begin{abstract}


   In this paper we propose a quiver model of matrix
   quantum mechanics with $8$ supercharges which, on
   a Higgs branch, deconstructs the worldsheet of
   Matrix String Theory. This discrete model evades
   the fermion doubling problem and, in the continuum
   limit, enhances the number of supersymmetries to
   sixteen. Our model is motivated by orbifolding
   the Matrix Model, and the deconstruction {\it 
   ansatz} exhibits a duality between target space
   compactification and worldsheet deconstruction.

  \end{abstract}

  \section{Introduction}

   \label{Sect1}

   Despite many great progresses in the last decade, a
   nonperturbative formulation of string theory, which
   is easy to work with, remains a hot topic of research.
   As alternatives to string field theory, a number of
   attempts have been tried by resorting to discrete 
   descriptions. Among them the string bit (or bit string) 
   model \cite{thorn} is conceptually appealing in that
   relativistic string is treated as a composite of point-like
   entities -- string bits. Recently there are revived
   interests \cite{bits} in string bit models, because
   the BMN correspondence \cite{bmn} between type-$IIB$
   string theory in a pp-wave background \cite{hull,matz}
   and $\natural = 4$ supersymmetric Yang-Mills (SYM) 
   theory in four dimensions is suggestive of a string 
   bit picture. However, since a string bit model 
   discretizes string worldsheet into a one-dimensional
   spatial lattice, depending on the way to do discretization,
   it may suffer from problems such as fermion doubling and/or
   absence of sufficient supersymmetries to ensure correct
   string interactions (for discussions of these problems,
   see \cite{thorn2,sochi}). In a sense, the BFSS Matrix Model
   \cite{BFSS} is in line with the spirit of string bits, but
   in the context of $11$-dimensional M-Theory. Nonetheless,
   within this framework nonperturbative type-$IIA$ strings,
   say, are described still by a two dimensional {\it field
   theory}, the Matrix String Theory \cite{dvv,motl,bs,hw},
   not by quantum mechanics.

   In this paper we propose a different approach to a discrete 
   formalism for nonperturbative string theory. We insist on 
   providing a (quantum) {\it mechanical} framework to describe 
   dynamics of string bits. Essentially what we propose is 
   to {\it deconstruct}, rather than to discretize,
   the string worldsheet. On one hand, it is known that
   to deconstruct a continuous dimension \cite{georgi} into a
   one-dimensional lattice, one needs to look up a quiver/moose
   field theory \cite{doug} with product gauge group and with
   bi-fundamental matter fields. Then assigning non-vanishing
   vacuum expectation value (VEV) to bi-fundamental bosonic
   fields gives rise to the kinetic energy (or hopping) as 
   well as gauge connections in the latticized direction.
   On the other hand, in certain cases deconstruction allows
   us to build lattice theories with exact supersymmetry free
   from the fermion doubling problem \cite{kaplan}. This is
   possible because a deconstructed lattice gauge theory
   allows at least some of the matter fields, in particular
   fermions, to live on {\sl links}. (Recall that in usual
   discretization, only the gauge fields live on links, 
   while matter fields, either bosonic or fermionic,
   all live on sites. For deconstruction (or large quiver 
   theories) in other contexts of string theory, see Refs. 
   \cite{halpern,ganor,skiba,motl2,mukhi,shiekh,kirsch}).
   We believe that the success of formulating superstring 
   theory in terms of discrete models would make 
   well-developed notions and techniques for one-dimensional 
   many-body models on a lattice, exactly solvable or not, 
   accessible to string/M theory.    

   Motivated by these ideas, we devise a supersymmetric
   quiver model of matrix quantum mechanics with 8
   supercharges, by orbifolding the BFSS Matrix Model
   in Section \ref{Sect2}. Then we show that on a Higgs
   branch it deconstructs string worldsheet with no
   fermion doubling (Section \ref{Sect3}). Moreover,
   it is shown in Section \ref{Sect4} that in the
   continuum limit the discrete model recovers the Matrix 
   String Theory, i.e. the two dimensional Super Yang-Mills 
   (SYM) theory that provides the DLCQ description of
   second quantized Type-$IIA$ superstrings. The Yang-Mills
   coupling is related to the string coupling in the just
   right manner as required by correct scaling for string
   interactions. A physical picture is given in Section
   \ref{Sect5} for our (de)construction of string world
   sheet in terms of the $D0$-brane description of Matrix
   Theory, which explicitly exhibits a duality between
   worldsheet deconstruction and target space compactification
   in M Theory. Finally we address the features and
   advantages of our quiver model for the dynamics of
   string bits (Section \ref{Sect6}). Appendix A is 
   devoted to some details of the transformation of 
   gamma martices and fermions at each stage of our 
   (de)construction procedure.

  \section{Orbifolding}

  \label{Sect2}

   To develop a quiver mechanics that describes the dynamics
   of ``string bits'' from deconstructing the
   worldsheet of Matrix String, we will start from the BFSS
   Matrix Theory, and consider it in an orbifold background
   $\complex^2/\integer_N$. The physical motivation is the
   following: We view the deconstructed closed string as a
   closed chain formed by $N$ beads and directed links (in
   both directions) connecting the neighboring beads. From
   the point of view of deconstruction, this closed chain
   is identical with the theory space (or quiver diagram
   \cite{doug}) with a discrete $\integer_N$ symmetry.
   With the dynamical degrees of freedom
   living on this space desirably being those of matrix
   mechanics, it is natural to start with the Matrix Theory
   that describes quantum mechanics of $D0$-branes. This
   leads to the idea of orbifolding the Matrix Theory on
   $\complex^2/\integer_N$. The reason to choose 
   $\complex^2$ is to make surviving supersymmetry as
   large as possible. The hope is that in a deconstruction
   phase and in a large-$N$ (continuum) limit, Matrix
   String Theory will be recovered. (For convenience,
   later we will call beads as sites, using the more
   familiar terminology of lattice theory.)

   Start with $U(KN)$ BFSS Hamiltonian, which reads
   (in temporal gauge $Y^0=0$ and in units of the
   $11$-dimensional Planck length $l_{p}$):
   \be
    {\cal H}= R\, {\rm Tr} \{
     {1\over 2}\Pi_I^2 - \frac{1}{4}[Y^I,Y^J]^2
     - \frac{i}{2} \theta^T \gamma^I [Y_I,\theta] \}.
   \label{BFSSaction}
   \ee
   Here $R$ is the radius of the compactified light cone;
   $Y^I$ ($I=1,\cdots,9$) are $KN$-by-$KN$ bosonic 
   hermitian matrices, whose superpartners are 16 
   fermionic hermitian matrices, $\theta$, which 
   transform as $SO(9)$ Majorana spinor. Moreover, 
   Tr is the trace over $KN$-by-$KN$ matrices. (The 
   transpose $T$ of $\theta$ acts only on the spinor
   index $\alpha=1,\cdots,16$, which we have suppressed.)

   Now we choose $\complex^2$ to be the hyperplane with
   $I=6,7,8,9$, and define
   \be
    Z^1 = {1\over\sqrt{2}}(Y^6 + iY^7), \qquad
    Z^2 ={1\over \sqrt{2}}(Y^8 - iY^9).
   \label{OC}
   \ee
   The orbifolding conditions by the action of $\integer_N$ read
   \bea
   \nonumber
    {\bf U}^\dag Y^I {\bf U} &=& [\omega^{M_{89}-M_{67}}]^I_J Y^J,\\
    {\bf U}^\dag \theta^\alpha {\bf U} &=& [\omega^{\sigma_{89}
                -\sigma_{67}}]^\alpha_\beta \theta^\beta.
   \label{ORB}
   \eea
   Here the embedding of $\integer_N$ into $U(KN)$ is
   given by ${\bf U} = U \otimes \unit_K$ with the
   $N$-by-$N$ clock matrix $U=diag (\omega,\omega^2,\ldots,
   \omega^N)$ and $\omega = \exp (i2\pi/N)$. $M_{67}$,
   $M_{89}$, $\sigma_{67}$ and $\sigma_{89}$ are rotation
   generators in the $(6,7)$-plane and $(8,9)$-plane for
   a vector and a spinor, respectively.

   By now it is well-known that with $Y^I$ and $\theta$
   written as $N$-by-$N$ block-matrices with each block
   being $K$-by-$K$, the above orbifolding conditions
   (\ref{ORB}) force many blocks to vanish. The outcome
   is a quiver theory with a lattice interpretation.
   More concretely, $Y^i$ ($i=1,2,3,4,5$) are 
   block-diagonal; the non-vanishing diagonal blocks 
   are the variables living on {\sl sites} of the circular 
   quiver diagram. $Z^a$ ($a=1,2$) have non-vanishing 
   blocks only just above each diagonal block and at
   the left-bottom corner; these non-vanishing blocks, as
   well as their hermitian conjugate, are variables live
   on the {\sl directed links} connecting the neighboring
   sites in the quiver diagram. Similarly for fermions, after
   a change of spinor basis, the 16 components of $\theta$ 
   are sorted into two groups: eight of them, $\Psi^\alpha$ 
   ($\alpha=1,\cdots,8$), live on sites, while the remaining 
   eight, $\tld{\Psi}^\beta$ ($\beta=1,\cdots,8$), live on 
   links connecting neighboring sites. (Their expressions in 
   terms of the origianl $\theta$ are given in Appendix.)

   Therefore, after orbifolding, the BFSS Hamiltonian
   on $\complex^2/\integer_N$ is of the following form:
   \bea
    \nonumber
    {\cal H} &=& R\, {\rm Tr} \{
     {1\over 2}\Pi_i^2 + \Pi_a\Pi_a^\dag 
     - \frac{i}{2} (\Psi^\dag \sigma^i [Y_i,\Psi] + \tld{\Psi}^\dag
     \bar{\sigma}^i [Y_i,\tld{\Psi}]) \\  \nonumber
     &&
     - \frac{i}{\sqrt{2}}(\Psi^\dag \sigma^a_{-} [Z_a,\tld{\Psi}]
     + \Psi^\dag \sigma^a_{+} [Z_a^\dag,\tld{\Psi}]
     + \tld{\Psi}^\dag \sigma^{a\dag}_{+} [Z_a,\Psi]
     + \tld{\Psi}^\dag \sigma^{a\dag}_{-} [Z_a^\dag,\Psi]) \\
     \nonumber
     &&
     - \frac{1}{4}[Y^i,Y^j]^2 - [Y^i,Z^a][Y^i,Z^{a\dag}] \\
     &&
     + {1\over 2}[Z^a,Z^{a\dag}]^2 - [Z^1,Z^2][Z^{1\dag},Z^{2\dag}]
     - {1\over 2}[Z^1,Z^{2\dag}][Z^{1\dag},Z^2]   \}.
    \label{action}
   \eea
   A representation of the 8-by-8 $\sigma$- and $\bar{\sigma}$-matrices, 
   which are deduced from the gamma matrices in Eq.~(\ref{BFSSaction}), 
   is given in Appendix.

   If we had not taken the temporal gauge, the gauge potential
   $Y^0$ in the BFSS Matrix Theory should be orbifolded too,
   turning into a site variable in accordance with Eq. (\ref{ORB})
   with $I=0$. The Lagrangian of the orbifolded system
   (\ref{action}), with $Y^0$ recovered, has an unbroken
   gauge group $U(K)^N$, one factor at each site, which is 
   the block-diagonal subgroup of the original $U(KN)$. The 
   site variables $Y^0$, $Y^i$ and $\Psi^\alpha$
   transform as the adjoint representation under the unbroken
   gauge group at each site. As for link variables $Z^a$ and
   $\tld{\Psi}^\beta$, they transform as the bi-fundamental
   representation of the two $U(K)$'s at neighboring sites
   connected by the link. In each term of Eq.~(\ref{action}),
   there is no ambiguity for what site indices should be taken
   for each factor: the unbroken gauge invariance dictates the
   site indices of each factor.

   Our system (\ref{action}) has a supersymmetry with $8$ 
   supercharges parameterized by a site spinor $\epsilon$:
   \[
    \delta_\epsilon Y_0 = \epsilon^\dag \Psi, \quad
    \delta_\epsilon Y^i = \epsilon^\dag \sigma^i \Psi, \quad
    \delta_\epsilon Y^m = 2\epsilon^\dag \sigma^m \tld{\Psi},
   \]
   \[
    \delta_\epsilon \Psi =
    -2i \sigma^{0i}[\nabla_0,Y_i]\epsilon
    - \sigma^{ij}[Y_i,Y_j]\epsilon - \sigma^{mn}[Y_m,Y_n]\epsilon,
   \]
   \be
    \delta_\epsilon \tld{\Psi} =
    -2i \sigma^{0m}[\nabla_0,Y_m]\epsilon
    - 2\sigma^{im}[Y_i,Y_m]\epsilon,
   \ee
   where $\nabla_0 =\partial/(R\partial t) -i [Y^0,\cdot]/2\pi$.

 \section{Deconstructing}
 \label{Sect3}

   The above orbifolded action (\ref{action}) does not contain
   hopping terms, i.e. no bilinear terms involving neighboring
   sites on the lattice. To generate desired string
   oscillations, we need to introduce hopping terms by carrying
   out another key step in the procedure of deconstruction:
   to consider the Higgs branch with non-vanishing VEV for
   bosonic link variables. The necessity to do so is natural
   from the Matrix Theory point of view: After orbifolding,
   the Hamiltonian ~(\ref{action}) describes K $D0$-branes 
   at the orbifold singularity, i.e. at the origin of 
   $\complex^2$. We need to pull these $D0$-branes away from
   the singularity, by giving the variables $Z^a$ ($a=1,2$) 
   a non-vanishing VEV.

   Let us reparameterize $Z^a$ ($a=1,2$) in terms of polar
   coordinates:
   \be
    \label{theta}
     Z^1 = (\bra Z^1 \ket + \rho_1) e^{i(\vartheta + \varphi)},
     Z^2 = (\bra Z^2 \ket + \rho_2) e^{i(\vartheta - \varphi)}.
    \ee
   The orbifolding conditions (\ref{ORB}) are then equivalent to
   \be
   \label{ORBE}
    \vartheta \sim \vartheta + \frac{2\pi}{N},
   \ee
   keeping other variables unaffected. (Thus the action of the
   $\integer_N$ symmetry is implemented as a discrete subgroup of 
   translations in the angular $\vartheta$-direction.)

   For convenience, we choose the following moduli conditions
   for the VEV of bosonic link variables:
   \be
   \label{VEV}
    \bra Z^1 \ket = - \frac{N\, R_9} {\sqrt{2}} {\bf V},
    \qquad \bra Z^2 \ket = 0.
   \ee
   (A choice of the classical moduli different from Eq. (\ref{VEV})
   is expected to differ merely by irrelevant operators in the
   continuum limit, in the sense of renormalization group when
   flowing to the infrared region.)  Here the minus sign on the
   right-hand side is a convention. $R_9$ is the length scale
   introduced due to the VEV of $Z^1$; its physical meaning is
   the radius of the M-circle in Matrix String theory via the
   ``9-11 flip'', which is related to the $IIA$ coupling by
   $R_9=g_sl_s$. ${\bf V}$ is the shift matrix given by
   $V\otimes \unit_K$,
   \[
   V:=
     \left(
      \begin{array}{ccccc}
       0&1&0&\cdots&0\\
       0&0&1&\cdots&0\\
       &&&\ddots &\\
       0&0&0&\cdots&1\\
       1&0&0&\cdots&0
      \end{array}
     \right).
   \]
   Then we expand the variables around their VEV:
   \be
   \label{expand}
   Z^1= \bra Z^1\ket + \frac{1}{\sqrt 2} (X^6+iA_1) , 
   \qquad Z^2= \frac{1}{\sqrt 2} (X^7-iX^8),
   \ee
   and substitute these expansions into the orbifolded 
   Hamiltonian ~(\ref{action}), we get a quiver model 
   of matrix mechanics, which will be interpreted as a 
   model that desconstructs worldsheet of Matrix String 
   Theory.

   To facilitate comparison with the conventions used
   in Matrix String Theory, we specify the trace to be
   \be
    {\rm Tr}=\frac{1}{NR_9}\sum\limits_{n=1}^N {\rm tr}
   \label{rescale}
   \ee
   where tr is the trace over $K$-by-$K$ matrices,
   and $n$ labels the sites in the quiver diagram.
   Then the quiver model can be viewed as a lattice
   Hamiltonian, with a lattice constant determined by
   the inverse VEV of $Z^1$,
   \be
    a= \frac{2\pi}{NR_9}.
   \label{A}
   \ee
   In this way, the overall factor $1/N\,R_9$ in Eq. 
   (\ref{rescale}) essentially plays the same role 
   as the lattice constant usually in front of the
   summation over sites in a discretized Hamiltonian. 
   More concretely, one may understand the factor 
   $1/N$ as coming from orbifolding, while the factor 
   $1/R_9$ later will be seen to control the size of 
   string worldsheet.

   One of the advantages of our model is that, as a 
   discrete model, it evades the problem of fermion 
   doubling. To show this, let us consider the spectrum 
   of fermions in our model. Write $\Psi=(\chi^+,\chi^-)^T$,
   $\tld{\Psi}=(\eta^-,\eta^+)^T$. The free fermionic 
   part of the Hamiltonian reads
   \bea
    \nonumber
    {\cal H}_{f.f.} &=& \frac{R}{4\pi}a\sum\limits_{n=1}^N {\rm tr} 
     \{ \chi^{+\dag}_n \alpha^1 \frac{\eta^-_n - \eta^-_{n-1}}{a}
     +\eta^{-\dag}_n \alpha^1 \frac{\chi^+_{n+1}-\chi^+_n}{a} \\
     &&
     -\chi_n^{-\dag} \alpha^1 \frac{\eta^+_n - \eta^+_{n-1}}{a}
     -\eta^{+\dag}_n \alpha^1 \frac{\chi^-_{n+1}-\chi^-_n}{a}
    \}.
   \label{SFF}
   \eea
   Here we use the convention that fermion at site $n$ is 
   denoted as $\chi^\pm_{n}$, while fermions on the link 
   connecting sites $n$ and $n+1$ as $\eta^\pm_{n}$), and 
   $\alpha^1 = \unit_2\otimes\tau_2$
   with $\tau_2$ the usual 2-by-2 Pauli matrix. We observe
   that the Hamiltonian (\ref{SFF}) is just that for Susskind's
   staggered fermions \cite{SS}. Therefore as is well-known,
   there is no fermion doubling in the present
   context. The natural emergence of staggered fermions is
   due to the existence of link fermions $\tld{\Psi}$ in
   addition to the site fermions $\Psi$; they combine
   together to form staggered fermions. We note that link
   fermions were not considered before in usual lattice field
   theory models. Their appearance is due to the lattice
   interpretation of the quiver diagram resulting from
   orbifolding \cite{kaplan}.

  \section{Constructing}
   \label{Sect4}

   Now we proceed to consider the continuum limit of the
   quiver model that was developed in the last two sections.
   The continuum limit is defined to be $N\rightarrow
   \infty$, $a\rightarrow 0$ with $Na=2\pi/R_9$ fixed.
   In this limit, the distinction between site and link
   variables disappears, and the quiver diagram (or the
   theory space) turns into a continuous circle, which
   combines with the light cone time $t$ to form closed
   string worldsheet.

   The fixed value of $Na$ is now understood as the 
   worldsheet length. Let us define the string worldsheet
   coordinate $\sigma$ such that its range is $[0,2\pi]$,
   after rescaling the variables as
   \bea
      Y^i \rightarrow \frac{X^i}{\sqrt{R_9}}, 
     \quad (i=1,\cdots,5), \qquad
      X^m \rightarrow \frac{X^m}{\sqrt{R_9}}, 
     \qquad (m=6,7,8),
    \label{scale2}
   \eea    
   in accordance with the conventions used in the Matrix String 
   Theory \cite{dvv}. Moreover, to cast the kinetic term for
   the gauge field in the canonical form, we rescale $A_0=Y_0/R_9$ 
   and $A_1\rightarrow R_9\,A_1$. Adopting the normalization in
   which the total light-cone momentum $P_+\equiv K/R=1$, it is 
   straightforward to show that the orbifolded Hamiltonian 
   ~(\ref{action}), with the help of Eqs. (\ref{VEV}), (\ref{expand}), 
   (\ref{rescale}) and (\ref{scale2}), in the continuum limit 
   reproduces a $d=1+1$, $\natural=8$ SYM theory, whose action 
   is (in string with $l_s=1$)
   \bea
   \nonumber
    S_{C.L.} &=& \frac{1}{2\pi}\int d\tau \int\limits^{2\pi}_0
     d\sigma\, {\rm tr} \{ - {1\over 2} f_{\mu\nu}f^{\mu\nu}
     - {1\over 2}[\nabla_\mu, X^I][\nabla^\mu, X^I]
     + \frac{g_{YM}^2}{4}[X^I,X^J]^2 \\
     &&
     + \psi^T [\nabla_0,\psi] - \psi^T \tau_3 [\nabla_1, \psi]
     + ig_{YM} \psi^T \beta^I [X_I,\psi] \}.
   \label{SCL}
   \eea 
   Here the worldsheet indices $\mu,\nu =0, 1$, the target
   space index $I$ now runs from $1$ to $8$, and
   \[
    f_{\mu\nu} = \partial_\mu A_\nu - \partial_\nu A_\mu
    -ig_{YM}[A_\mu,A_\nu],
   \]
   \[
    [\nabla_\mu,X^I] = \partial_\mu X^I -
    ig_{YM}[A_\mu,X^I].
   \]
   (The overall factor $R$, the light cone radius in DLCQ Matrix 
   Theory, has been absorbed into the definition of the worldsheet 
   time, together with a factor of the inverse $R_9$ into $\sigma$.)

   In Eq.~(\ref{SCL}), we have diagonalized the covariant
   kinetic energy terms of the fermions $(\chi^\pm,\eta^\pm)$
   in terms of a new fermion field $\psi$ and separated the 
   two-dimensional left- and right-moving spinors. The details 
   of the transformations and the resulting $\beta^I$-matrices 
   in the Yukawa coupling terms can be found in Appendix. 
   Now it is easy to check that the continuum limit recovers 
   the rotational symmetry for the eight transverse coordinates
   $X^I$, and the left- and right-moving parts of $\psi$ 
   transform as ${\bf 8_s}$ and ${\bf 8_c}$, respectively.

   Thus the action (\ref{SCL}) is exactly that of the SYM
   obtained before by compactifying Matrix Theory on a 
   circle \cite{taylor,dvv,motl,bs,hw}, which describes 
   type-$IIA$ superstrings (or type-$IIB$ $D1$-strings
   \cite{hw}). Moreover, the Yang-Mills coupling $g_{YM}$ 
   is related to the type-$IIA$ string coupling $g_s$ by
   \be
   g_{YM}^2=\frac{1}{g_s^2 l_s^2},
   \label{coupl}
   \ee
   exactly the same as that in the Matrix String Theory, 
   where this relation indicates that the string coupling 
   $g_s$ scales inversely with worldsheet length \cite{dvv}. 
   From the M theory point of view, this means that the 
   Yang-Mills coupling should be inversely proportional to 
   the M-circle radius $R_9$, to give correct scaling for 
   string interactions.

   Extended supersymmetry is enhanced to $16$ supercharges
   in the action (\ref{SCL}). (Similar enhancement of
   supersymmetries has been found in deconstructed models
   before, say in Refs. \cite{kaplan,motl2}). Actually the
   above action is precisely the dimensional reduction of
   $d=1+9$, $\natural=1$, $U(K)$ SYM to $d=1+1$; the
   existence of $16$ supersymmetries is evident.

  \section{M Theory Picture}

   \label{Sect5}

  The deconstruction procedure we have presented in above
  sections has an explicit physical picture in M theory.
  The orbifolding in Section \ref{Sect2} (see Eq.
  (\ref{ORB})) is actually to gauge the discrete translations
  in the angular $\vartheta$-direction, defined in Eq.
  (\ref{theta}), as manifestly shown by Eq. (\ref{ORBE}).
  The orbifolded Hamiltonian ~(\ref{action}) describes $K$
  $D0$-branes at the $\complex^2/\integer_N$ orbifold
  singularity. The assignment in Eq. (\ref{VEV}) of
  non-vanishing VEV to the link variables $Z^1$
  means that the classical configuration of $D0$-branes
  is now away from the orbifold singularity by a
  distance $N R_9$ in the wedge $0\le \vartheta \le
  2\pi/N$. With the two edges identified in the 
  orbifold picture, the $D0$-branes are now in 
  a very thin cone and far away from the tip, so 
  essentially they can be considered to live on a {\it 
  cylindrical geometry}, whose transverse circle is of 
  radius $R_9$. What one has achieved is the compactification 
  of M theory on a circle; our choice of the moduli 
  implies that it is in the $7$th direction. This is 
  the so-called M-circle that relates M theory to 
  type-$IIA$ string theory. Note that the identifcation 
  of the VEV of bosonic link variables with a 
  compactifcation radius makes sense {\it only} in 
  string/M theory; in usual deconstruction in 
  non-string-theory context, it is the inverse VEV 
  that is proportional to the lattice constant in the 
  (de)constructed extra dimension \cite{georgi,kaplan}.

  The worldsheet in Matrix String Theory is known 
  to live on the dual circle of the M-circle. To 
  (de)construct it, we have to assign to the 
  circular quiver diagram a lattice constant $a$, 
  such that the circumference is just that of the 
  dual circle, $2\pi/R_9$. The lattice constant 
  $a$ given by the deconstruction {\it ansatz} 
  (\ref{A}) is just right. Moreover, this choice 
  of $a$ also turns the coordinate $Y^7$ in 
  Eq. (\ref{expand}) into a discretized version 
  of a covariant derivative, just as the 
  compactification in M theory required. So in the 
  M theory picture the Matrix String worldsheet is 
  of radius $1/R_9$. (In the conventions for Matrix 
  String Theory, the lattice constant is essentially 
  $2\pi/N$ with the normalization $Na=2\pi$.) 
  Therefore, what the quiver diagram deconstructs 
  is indeed the Matrix String worldsheet, and the 
  deconstructed lattice constant (\ref{A}) exhibits 
  explicitly a duality between worldsheet 
  (de)construction and target space compactification 
  in M theory. Previously in Matrix Theory this 
  duality appears as an {\it ansatz} for solving 
  the quotient conditions for compactification. 
  Here in this note the duality naturally appears 
  due to deconstruction of worldsheet, which also 
  gives rise to correct scaling for string 
  interactions.

  \section{Discussions}

  \label{Sect6}

   In this note, we have proposed to deconstruct string
   worldsheet, resulting in a quiver model of supersymmetric
   matrix quantum mechanics, which provides the DLCQ description
   of type-$IIA$ superstring bits. The Matrix String Theory
   is recovered exactly in the continuum limit. In the spirit 
   of providing a lattice formulation of string theory, our 
   proposal is similar to string bit models. But there are 
   several important differences arising from deconstruction 
   in contrast to naively discretizing the string action.

   First, in addition to gauge fields, there are other dynamical
   variables living on links. On one hand, there are link
   fermions. This makes it possible to evade the famous fermion
   doubling problem in the usual lattice theory. Also the fact 
   that the theory has eight extended supersymmetry which is 
   enhanced in the continuum limit to sixteen supercharges is 
   closely related to the existence of link fermions. On the 
   other hand, unlike the Wilson lines in lattice gauge theory, 
   the bosonic link variables are non-unitary. Since the 
   geometric significance of link variables is parallel transport,
   the non-unitary portion in them implies a dynamical effects
   on geometry. We believe the introduction of the non-unitary
   links should have profound effects on quantum gravity, hence on
   string/M theory as well as on holography.

   Secondly, the lattice spacing $a$ is now related to the
   expectation value of bosonic link variables and, therefore,
   becomes dynamical in the context of deconstruction, as a 
   characteristics shared by all deconstruction models. This 
   is a wonderful feature, distinct from usual discretization 
   in which lattice spacing is introduced by hand merely as a 
   means of regularization. In other words, the lattice 
   spacing may ultimately be a physical quantity determined 
   by the underlying nonperturbative dynamics, while the 
   continuum geometry emerges effectively in the infrared 
   regime. It is in this sense we think the quiver matrix 
   quantum mechanics may perhaps be more fundamental/microscopic 
   than Matrix String Theory is.

   Normally the idea of deconstruction is used to deal
   with a higher dimensional {\it target space} in terms of a
   lower dimensional theory. In this note what we have attempted
   to do is to deconstruct {\it string worldsheet}. The success
   seems to imply a {\sl worldsheet/target-space duality}. Now
   we have two different ways to obtain the Matrix String
   Theory. The standard routine is to compactify the BFSS matrix
   quantum mechanics on a circle and obtain the Matrix String
   Theory as an SYM on the dual circle. In this note,
   we orbifolded the BFSS model, and consider the deconstruction
   phase of the resulting quantum mechanics; after taking the
   continuum limit, we recover Matrix String Theory. So what
   we have done is actually the deconstruction of the dual space
   of the compactified circle in Matrix Theory. In this way,
   one may say that we have a duality between worldsheet
   deconstruction and target space compactification in string/M
   theory. The well-known $IIA/M$ duality can be viewed as a 
   prototype of this duality; namely the eleventh dimension in 
   M theory is originated from trading the worldsheet spatial 
   dimension to target space.

   Our quiver mechanics model of strings is motivated 
   by orbifolding the BFSS Matrix model and taking 
   the deconstruction phase. However this proposal 
   can be taken directly as the starting point of an 
   alternative to string field theory. The hope is that 
   the quiver matrix mechanics (with finite $N$) could 
   be shown to be mathematically more manipulable and 
   less singular than string field theory. One further 
   advantage of studying the quiver mechanics seems to 
   be that it opens the door for numerical study of 
   string theory. One can start with the $K=1$ case, 
   which provides a well-behaved, discretized toy model for 
   superstring theory. Another possible advantage of the 
   quiver mechanical approach is to provide a framework 
   for studying renormalization group flow or even phase 
   diagrams or phase transitions in string/M theory. 

   In this note, we have studied the simplest case of quiver 
   mechanical models in string theory. It is obvious that 
   more complicated quiver mechanical models can be 
   constructed for higher dimensional extended objects, e.g. 
   $Dp$-branes, or for strings in other backgrounds, such as 
   pp-waves and some noncommutative geometric backgrounds. 
   It is expected that the success of formulating superstring 
   theory in terms of discrete models should make well-developed 
   notions and techniques for one- or two-dimensional many-body 
   models on a lattice, exactly solvable or not, accessible to 
   string/M theory.   
        
   {\bf Acknowledgements} JD thanks the High Energy Astrophysics 
   Institute and Department of Physics in the University of 
   Utah, and Profs. K. Becker, C. DeTar and D. Kieda for warm 
   hospitality and financial support.  
 
   {\it Note Added} When this work was completed and being 
   written up, a preprint (hep-th/0306147) by Danielsson {\it 
   et al.} appeared \cite{daniel}, in which the use of staggered 
   fermions in a string bit model was also proposed to avoid the 
   fermion doubling problem.

   \appendix

  \section{Representations of Gamma Matrices and Fermions}

   In this Appendix, we give some technical detalis 
   about how fermions are transformed at each step of
   our (de)construction. There are good reasons to 
   believe that the sixteen fermions in the continuum 
   action ({\ref{SCL}) must possess the correct chirality 
   properties under the transverse $SO(8)$ rotations to
   describe type-$IIA$ strings. However, it is non-trivial
   to see how this happens explicitly, since the $SO(8)$ 
   symmetry is violated by orbifolding and gets recovered 
   only in the continuum limit. This is also crucial for 
   the supersymmetries broken by orbifolding to get 
   recovered. 
   
   In Eq. (\ref{BFSSaction}) the fermionic coordinates
   $\theta$ is an $SO(9)$ Majorana spinor of 16 real
   components, which can be labelled by four indices
   $s_1,s_2,s_3,s_4$, each taking two values only. 
   The gamma matrices $\gamma^I$ ($I=1,\cdots,9$) 
   can be read off from the standard $\Gamma$-matrices 
   in ten dimensions \cite{GSW}, and are represented 
   by a direct product of four $2$-by$2$ matrices:
   \bea
     \gamma^1 = \epsilon \otimes \epsilon \otimes \epsilon \otimes \epsilon
     &,& 
     \gamma^2 = \tau_1 \otimes \unit \otimes \epsilon \otimes \epsilon,
 \nonumber  \\  
     \gamma^3 = \tau_3 \otimes \unit \otimes \epsilon \otimes \epsilon &,&
     \gamma^4 = \epsilon \otimes \tau_1 \otimes \unit \otimes \epsilon, 
 \nonumber  \\
     \gamma^5 = \epsilon \otimes \tau_3 \otimes \unit \otimes \epsilon &,& 
     \gamma^6 = \unit \otimes \epsilon \otimes \tau_1 \otimes \epsilon,
 \nonumber  \\
     \gamma^7 = \unit \otimes \epsilon \otimes \tau_3 \otimes \epsilon &,&
     \gamma^8 = \unit \otimes \unit \otimes \unit \otimes \tau_1,
 \nonumber  \\
     \gamma^9 = \unit \otimes \unit \otimes \unit \otimes \tau_3 &.&
   \eea
   Here we used $\unit$ and $\tau_i$ to denote the 2-by-2 unit 
   and Pauli matrices; and $\epsilon=i\tau_2$. 

   The change from $\theta$-spinors to the site fermions 
   $\Psi$ and the link fermions $\tld{\Psi}$ in Eq. 
   (\ref{action}) only involves the indices $s_3$, $s_4$:
   \be
   \label{Trans}
    \left(
     \begin{array}{c}
      \theta^{s_1s_2++} \\
      \theta^{s_1s_2+-} \\
      \theta^{s_1s_2-+} \\
      \theta^{s_1s_2--}
     \end{array}
    \right)
    ={1\over 2}
    \left(
     \begin{array}{c}
      (\theta^{s_1s_211}+i\theta^{s_1s_212})+i(\theta^{s_1s_221}
        +i\theta^{s_1s_222}) \\
      (\theta^{s_1s_211}-i\theta^{s_1s_212})+i(\theta^{s_1s_221}
        -i\theta^{s_1s_222}) \\
      (\theta^{s_1s_211}+i\theta^{s_1s_212})-i(\theta^{s_1s_221}
        +i\theta^{s_1s_222}) \\
      (\theta^{s_1s_211}-i\theta^{s_1s_212})-i(\theta^{s_1s_221}
        -i\theta^{s_1s_222})
     \end{array}
    \right),
   \ee
   Then we have $\Psi=(\theta^{s_1s_2++},\theta^{s_1s_2--})^T$,
   $\tld{\Psi}=(\theta^{s_1s_2-+},\theta^{s_1s_2+-})^T$. 

   This transformation gives rise to the following matrices 
   in Eq. ({\ref{action}):
   \bea
    \nonumber
     \sigma^1 = \tau_2 \otimes \tau_2 \otimes \unit 
          &,& \bar{\sigma}^1 = -\sigma^1, \\
    \nonumber
     \sigma^2 = -\tau_1 \otimes \unit \otimes \unit 
          &,& \bar{\sigma}^2 = -\sigma^2, \\
    \nonumber
     \sigma^3 = -\tau_3 \otimes \unit \otimes \unit 
          &,& \bar{\sigma}^3 = -\sigma^3, \\
    \nonumber
     \sigma^4 = \tau_2 \otimes \tau_1 \otimes \tau_3 
          &,& \bar{\sigma}^4 = \sigma^4, \\
    \nonumber
     \sigma^5 = \tau_2 \otimes \tau_3 \otimes \tau_3 
          &,& \bar{\sigma}^5 = \sigma^5, \\
    \nonumber
     \sigma^6 = i\unit \otimes \tau_2 \otimes \unit &,&
     \sigma^7 = \unit \otimes \tau_2 \otimes \tau_3, \\
    \nonumber
     \sigma^8 = -\unit \otimes \unit \otimes \tau_2 &,&
     \sigma^9 = \unit \otimes \unit \otimes \tau_1. \\
     \sigma^1_\pm = \frac{\sigma^6 \pm i\sigma^7}{2} &,&
     \sigma^2_\pm = \frac{\sigma^8 \mp i\sigma^9}{2}.
    \eea
  
   In Eq. (\ref{SFF}) we have $\chi^+ = \theta^{s_1s_2++}$, 
   $\chi^- = \theta^{s_1s_2--}$, $\eta^+ = \theta^{s_1s_2+-}$, 
   $\eta^- = \theta^{s_1s_2-+}$. Hence $\chi^{-\dag} = \chi^+$, 
   $\eta^{-\dag} = \eta^+$.

   The continuum limit irons out the distinction between site
   and link fermions, and the fermionic part of the action reads
   \bea
    S &=& \frac{1}{2\pi} \int d^2\sigma\, {\rm tr} \{
     \psi^T [\nabla_0,\psi] - \psi^T \gamma^7 [\nabla_1, \psi] \\
     &&
     + ig_{YM} (\psi^T \gamma^i [X_i,\psi] 
     + \psi^T \gamma^8 [X_7,\psi] + \psi^T \gamma^9 [X_8,\psi] )\}.
   \eea
   Here $\psi$ is just the image of $\chi$ and $\eta$ under the
   inverse transformation of Eq.~(\ref{Trans}), with positions 
   of indices $s_2$ and $s_3$ interchanged only for convenience.
   To separate the left- and right-moving components we need to 
   do orthogonal transformations to diagonalize $\gamma^7$ to 
   be of the form $\unit\otimes\unit\otimes\unit\otimes\tau_3$.
   Observe that $\gamma^7 = \unit \otimes \epsilon \otimes 
   \tau_3 \otimes \epsilon$. We first try to do an orthogonal 
   transformation $N_1$ involving only the second and the fourth 
   factors in the product, such that
   \be
    (\epsilon\otimes\epsilon)N_1 = N_1(\unit\otimes\tau_3).
   \ee   
   Then we do a second orthogonal transformation $N_2$ 
   involving only the third and the fourth factors in the 
   product, such that
   \be
    (\tau_3\otimes\tau_3)N_2 = N_2(\unit\otimes\tau_3).
   \ee
   The explicit form of $N_1$ and $N_2$ are
   \be
    N_1 = {1\over 2}\left(
     \begin{array}{cccc}
      1&1&0&0 \\
      0&0&1&1 \\

      0&0&-1&1 \\
      1&-1&0&0
     \end{array}
    \right), \qquad \quad
    N_2 = \left(
     \begin{array}{cccc}
      1&0&0&0 \\
      0&1&0&0 \\
      0&0&0&1 \\
      0&0&1&0
     \end{array}
    \right).
   \ee

   In the continuum action (\ref{SCL}) we use the same notation $\psi$
   for the transformed fermions, with the gamma matrices other than 
   $\gamma^7$ changing into eight $\beta$-matrices, and with $\gamma^7$ 
   into $\beta^9$:
   \bea
    \nonumber
    \beta^1 = \epsilon \otimes \tau_1 \otimes \unit \otimes \epsilon &,&
    \beta^2 = \tau_1 \otimes \epsilon \otimes \epsilon \otimes \tau_1, \\
    \nonumber
    \beta^3 = \tau_3 \otimes \epsilon \otimes \epsilon \otimes \tau_1 &,&
    \beta^4 = -\epsilon \otimes \tau_3 \otimes \unit \otimes \epsilon, \\
    \nonumber
    \beta^5 = \epsilon \otimes \unit \otimes \epsilon \otimes \tau_1 &,&
    \beta^6 = \unit \otimes \epsilon \otimes \unit \otimes \epsilon, \\
    \beta^7 = \unit \otimes \unit \otimes \tau_1 \otimes \tau_1 &,&
    \beta^8 = \unit \otimes \unit \otimes \tau_3 \otimes \tau_1.
   \eea

   The explicit expression $\beta^9 = \unit\otimes\unit\otimes\unit
   \otimes\tau_3$ shows that the left- and right-moving components 
   of $\psi$ have opposite chirality under the transverse $SO(8)$ 
   rotations. So the action (\ref{SCL}) descirbes type-$IIA$ strings.

 \end{document}